\documentclass{elsart}
\usepackage{epsfig,float,amssymb}
\journal{Astroparticle Physics}

\begin{document}

\begin{frontmatter}

\title{Feasibility of a magnetic suspension for second generation Gravitational Wave interferometers}

\author[na]{Monica Varvella\corauthref{cor}},
\corauth[cor]{Corresponding author.}
\ead{varvella@na.infn.it}
\author[na]{Enrico Calloni},
\author[na]{Luciano Di Fiore},
\author[na]{Leopoldo Milano},
\author[lal]{Nicolas Arnaud}

\address[na]{Istituto Nazionale di Fisica Nucleare, Sez. di Napoli, Universit\'a degli Studi di Napoli "Federico II"(Na), Italy.}

\address[lal]{Laboratoire de l'Acc\'el\'erateur Lin\'eaire, CNRS-IN2P3 and Universit\'e Paris Sud,\\B.P. 34, B\^atiment 200, Campus d'Orsay, 91898 Orsay Cedex (France).\protect\\}

\begin{abstract}

This paper deals with the use of a magnetic levitation system as a part of a multi-stage seismic attenuator for gravitational wave interferometric antennas. The proposed configuration uses permanent magnets in attraction to balance the suspended weight, plus a closed loop position control to obtain a stable levitation. The system is analyzed using a MATLAB simulation code to compute the forces exerted by extended magnets. The validity of this model has been tested by a comparison with the experimental data from a levitated suspension prototype. 
 
\end{abstract}


\end{frontmatter}

%\baselineskip = 2\baselineskip 

\section{Introduction}
 
Magnetic levitation systems find several applications 
\cite{Brandt,Geim,Jayawant,Jiles} in physics and engineering, ranging from small force measurements to transportation systems. In this paper, we analyze the possibility of using magnetic levitation as a future upgrade of the sophisticated suspensions used in long baseline interferometric Gravitational Wave (GW) detectors like, for example, Virgo \cite{virgo}.

The detection of GW is one of the most challenging fields of contemporary physics \cite{Blair,Saulson} and interferometers appear to be the most promising detectors looking for the first direct evidence of GW. Presently, several long baseline antennas have already started operation \cite{ligo,tama,geo}, or are still under construction \cite{virgo,aciga}. Due to the extreme weakness of the GW signals, the main requirement for these detectors is the isolation of the optical components -- used as free-falling test masses -- from any kind of external disturbance. For this reason, the whole optical path (e.g. three km per arm for Virgo) is under ultra high vacuum and all the mirrors are suspended to complex vibration isolation systems, which strongly reduce the effect of seismic vibration, that is the main limiting noise below few Hz.

The idea of using stable levitation for the mirror suspensions of long baseline GW detectors has been proposed by Drever \cite{Drever} for magnetic levitation, and Giazotto \cite{Giazotto} for electrostatic levitation. The main advantage of these systems is the possibility to hold the mirror without direct contact; therefore, these configurations avoid some noise contributions due to the presence of suspension wires -- for example the thermal noise associated with violin modes.

In the case of magnetic levitation, the presence of big magnets directly attached to the mirrors would decrease the mechanical quality factor of the internal modes, thus increasing the associated thermal noise, and should make the mirror position directly sensitive to external electro-magnetic (EM) noise. On the other hand, using an electrostatic levitation system is safe from the point of view of the mirror quality factor.

To obtain a stable levitation, it is necessary to use a servo-loop to control the vertical position of the mirror. In this case, the noise of the position sensor should be low enough, in order not to spoil the sensitivity of the antenna. For example, for the Virgo interferometer a sensitivity of $10^{-17} \ \mathrm{m} / \sqrt{\mathrm{Hz}}$ at 10~Hz is expected. Assuming a (optimistic) vertical to horizontal coupling factor of 1\%, the maximum acceptable noise for the sensor is $10^{-15} \ \mathrm{m} / \sqrt{\mathrm{Hz}}$. This specification is quite severe, although not impossible, and requires an interferometric readout system.

All these questions must be carefully investigated before concluding that direct levitation of the mirrors can be actually applied to GW detectors. Another possibility, studied in this paper, is the usage of a levitation system as an intermediate stage of a Virgo-like multi-stage seismic attenuator.

\begin{figure}
 \epsfxsize=17cm
 \centerline{\epsffile{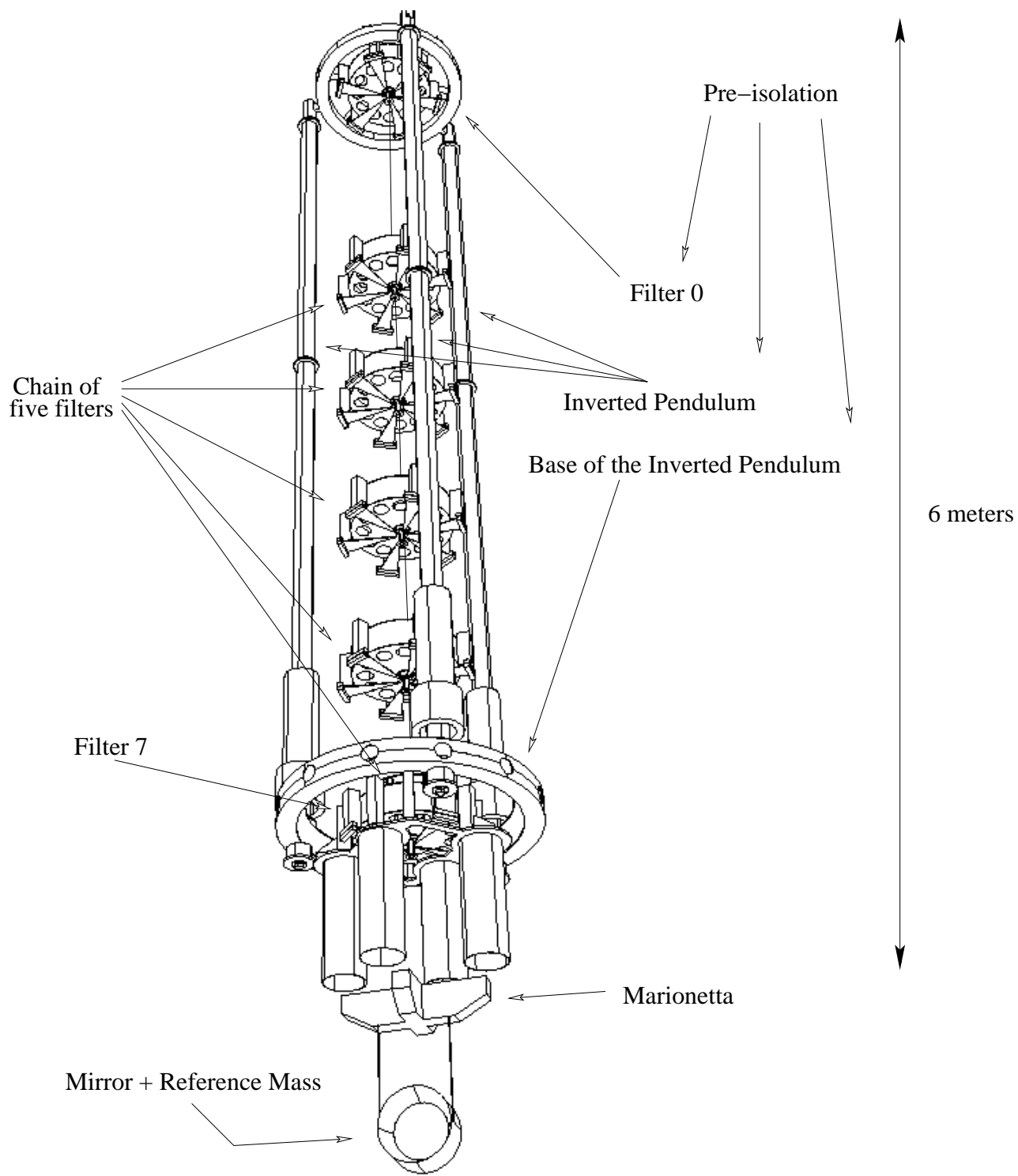}}
 \caption{\label{fig:f1} \sl{Simplified scheme of the Virgo Superattenuator.}}
\end{figure}

The Virgo seismic suspension, called Superattenuator (SA), is a pendulum with blade springs at each of its five stages, which provides a vibration attenuation in six degrees of freedom by 
more than 200~dB in the whole measurement frequency band of the antenna (4~Hz $\div$ 10~kHz). A detailed description of the SA can be found in \cite{SA}; in the following, we will only  review its most relevant features. Fig.~\ref{fig:f1} shows a Virgo SA which can be conceptually divided into two parts.

\begin{itemize} 
 \item The upper one, including the 6-meter inverted pendulum supporting the five attenuation stages.
 \item The lower one, called payload, made of an intermediate mass -- the {\sl{marionetta}} --, supporting the mirror itself, and of the {\sl{reference mass}}, a cylindrical shell suspended around the mirror with same mass and same barycenter. 
\end{itemize}

Tab.~\ref{tab:t1} lists the masses of the different SA components. The mirror orientation and position can be controlled using coils as actuators. They are either attached to the last stage of the upper part of the SA and acting on magnets glued on the marionetta, or fixed to the {\sl{reference mass}} and acting on small magnets directly fixed on the mirror edges. The lower part of the SA is in ultra-high vacuum ($< 10^{-7}$~mbar) to avoid contamination of the mirror surfaces by hydrocarbonates in particular, while the upper part is only under high vacuum ($10^{-6}$ mbar), because of the outgassing of several components of the suspension. To avoid the contamination of the mirrors, the two vacuum chambers are separated by a roof with a small conductance hole for the suspension wire.

Due to the low frequency thermal drifts of the system, the centering of the wire in conductance hole is quite difficult and requires periodic re-adjustment, or the use of a closed loop control for long term operation.

\begin{figure}
 \epsfxsize=12cm
 \centerline{\epsffile{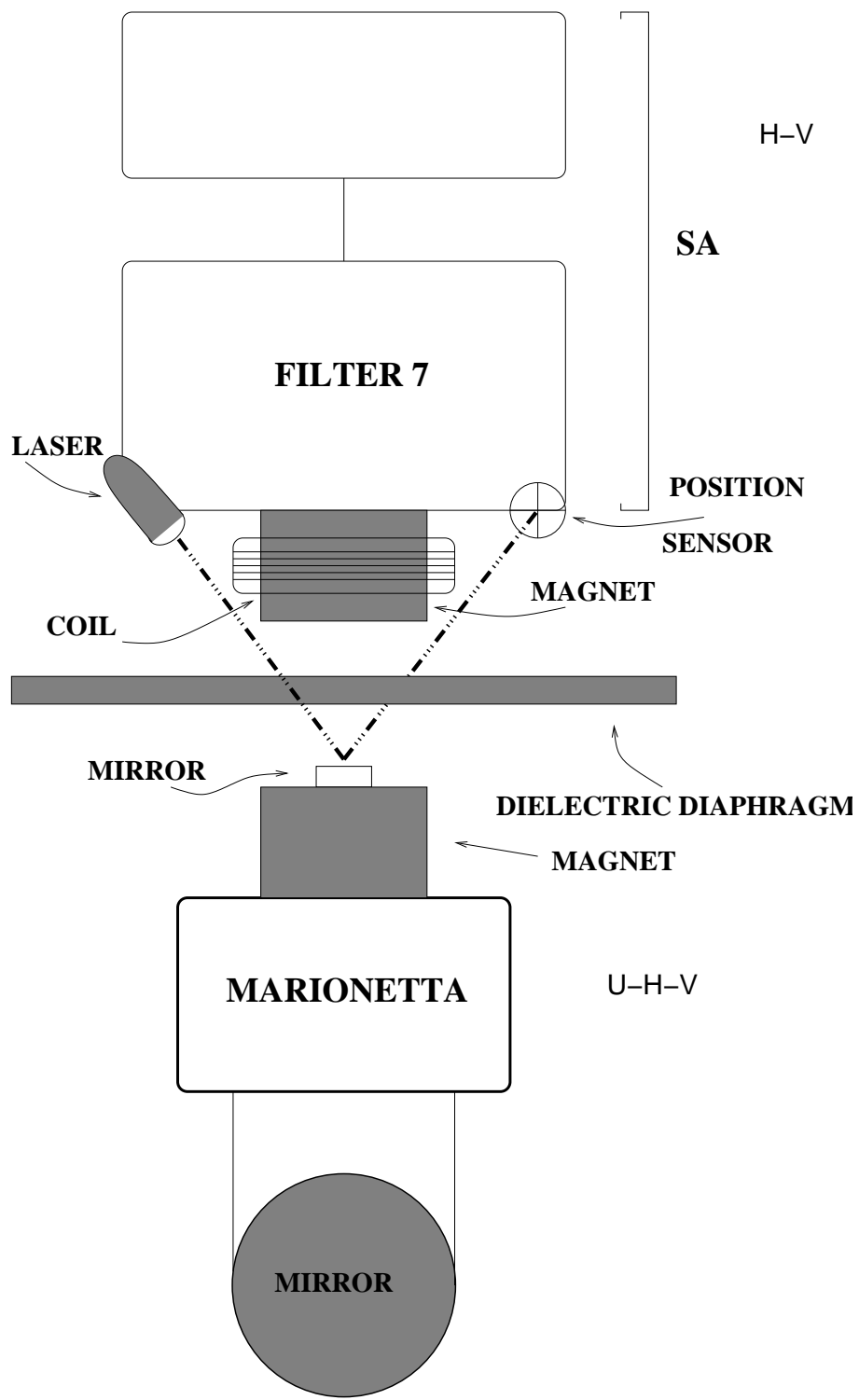}}
 \caption{\label{fig:f2} \sl{Possible scheme of the bottom part of a Superattenuator with a magnetic suspension.}}
\end{figure}

A possible alternative solution of the problem is the levitation of the {\sl{marionetta}}, as shown in Fig.~\ref{fig:f2}. In this way, the upper and lower parts of the vacuum chamber could be completely isolated by a dielectric and transparent roof, eliminating the conductance hole. This configuration is much less sensitive to the noise problems described earlier, as both the sensor noise and the external EM noise are injected at the level of the marionetta and are thus filtered by the last stage of the suspension, providing an additional attenuation of more than two orders of magnitude at 10 Hz. Indeed, the specification on the sensor noise can be relaxed to $10^{-13}~\mathrm{m} / \sqrt{\mathrm{Hz}}$.

The main goal of this work is thus to evaluate the possibility of suspending with magnetic levitation a given mass (on the order of the sum of the masses of the mirror and of the {\sl reference mass}) at manageable cost. 

In the following sections, we analyze the feasibility of such a levitated system. In Section \ref{sec:magnetic}, the general principle of stable levitation \cite{Jayawant} is shortly recalled. Then, the numerical procedures used to compute the forces between the different magnets are described. In Section \ref{subsec:simulation}, the simulation results for magnetic levitation applied to the Virgo suspensions are presented. %; advantages and drawbacks of the different possible solutions are then discussed in Section \ref{subsubsec:discussion}. 
An experimental test -- described in Section \ref{subsec:test} -- shows the feasibility of the magnetic levitation; the experimental results obtained on a small levitated suspension are compared with the predictions of some numerical simulations in order to check the validity of the model.
% Tab.~\ref{tab:t1} lists the masses of the different SA components.  

\section{Magnetic suspension system}
\label{sec:magnetic}

%\subsection{General considerations}

As it is well known, a stable levitation in a constant magnetic field can be obtained only with superconducting and diamagnetic materials 
\cite{Brandt,Geim,Jayawant,Jiles}. Using superconducting materials requires low temperature operation; thus, their integration in the seismic suspension of GW detectors looks complex and expensive. Moreover, for diamagnetic materials, levitating very small pieces requires intense fields, which makes the suspension of a payload of several tens of kilograms (see Tab.~\ref{tab:t1}) almost impossible. On the other hand, using a magnetic field of variable intensity controlled by a feedback system in association with permanent magnets and electromagnets is more promising.

\begin{figure}
 \epsfxsize=18cm
 \centerline{\epsffile{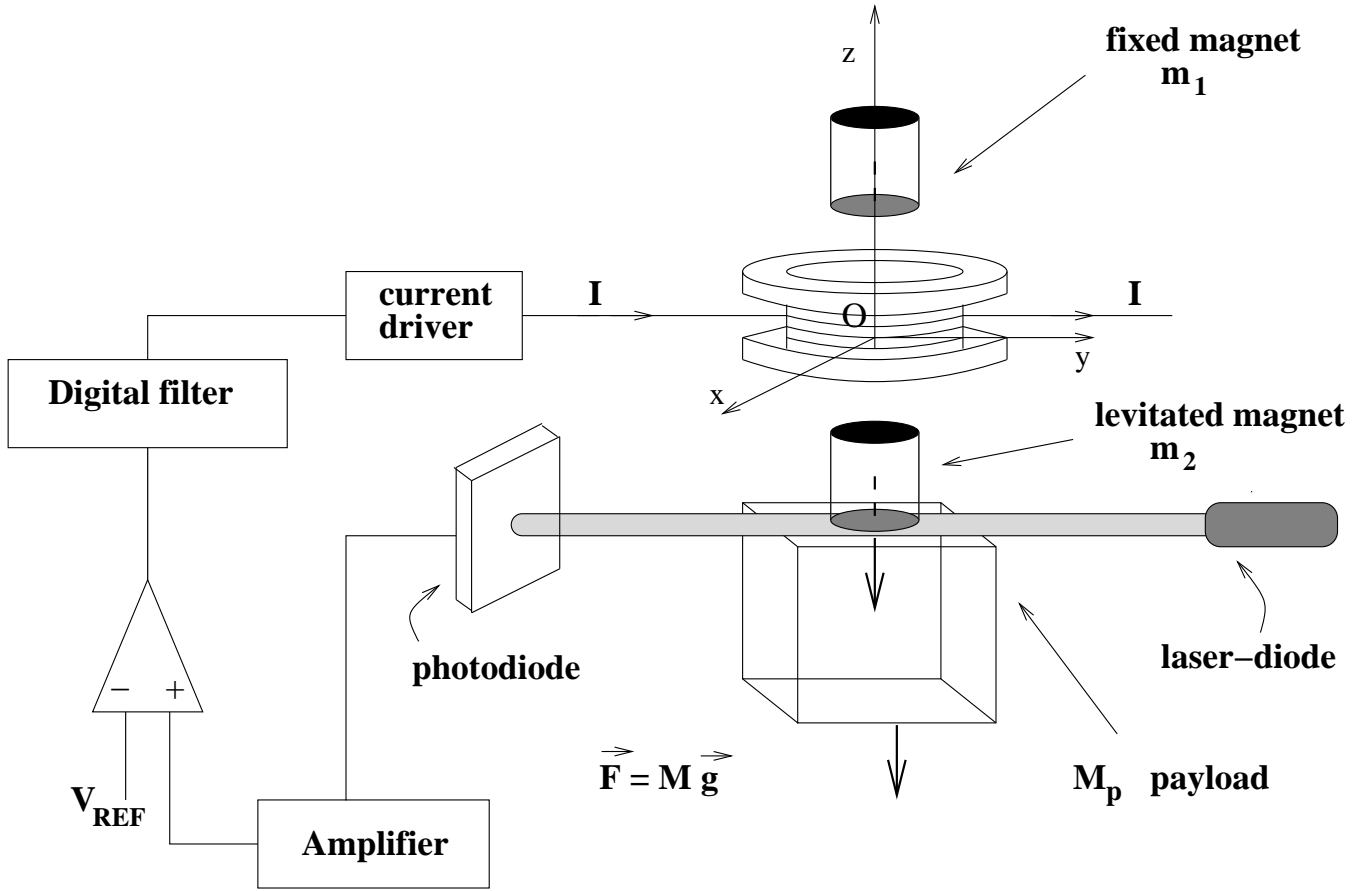}}
 \caption{\label{fig:f3} \sl{General principle of a magnetic suspension: a permanent magnet of mass $m_2$ is attached to the mass $M_p$ to be levitated (the payload in our case) and is attracted by a fixed magnet of mass $m_1$ which exactly balances the gravitational force. The equilibrium position is stable in the horizontal direction, but unstable in the vertical one; a stable configuration is obtained with the addition of a coil acting on the levitated magnet, with a current intensity $I$ depending on the vertical position of the levitated mass. The payload position is measured with any type of position sensor (e.g. a shadow meter composed by a photodiode and a laser-diode) and a feedback is digitally implemented to control the magnetic system. Note: if the fixed magnet $m_1$ is removed, a large DC current is necessary to balance the weight of the levitated piece.}} 
\end{figure}

The general principle of such device can be seen in Fig.~\ref{fig:f2} and \ref{fig:f3}: a permanent magnet of mass $m_2$ is attached to the mass to be levitated (for the Virgo SA, the payload and the mirror) and is attracted by a fixed magnet of mass $m_1$ (on the SA upper part) which exactly balances the gravitational force. In this configuration, the equilibrium position is stable in the horizontal direction, but unstable in the vertical one. This configuration can become stable if a coil acting on the levitated magnet is added, with a current intensity $I$ depending on the vertical position of the levitated mass. This position can be measured with any type of position sensor (a shadow meter for our experiment, see Section~\ref{subsec:test}). In principle, the fixed magnet $m_1$ could be removed, but in this case, a large DC current would be necessary to balance the weight of the levitated piece.

\subsection{The numerical simulation}
\label{subsec:numerical}

Testing the feasibility of such technique is not enough, as one cannot simply scale a system made of point-like magnets to a large device using extended magnets, such as the one needed for a Virgo SA: the dipole approximation is no more valid to compute the magnetic force. Therefore, an accurate calculation of the forces between the fixed and the levitated magnets is needed.

\begin{figure}
 \epsfxsize=18cm
 \centerline{\epsffile{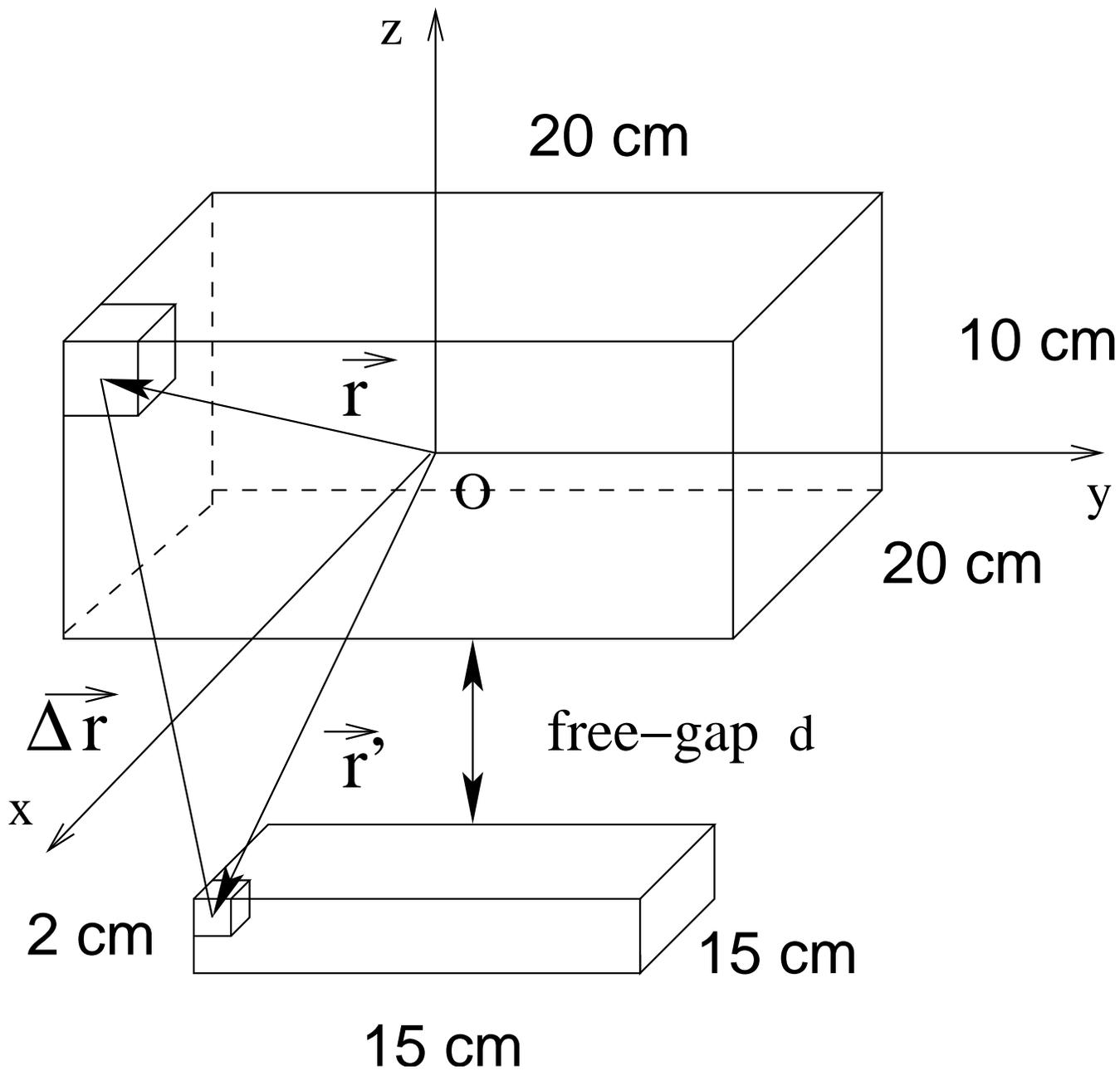}}
 \caption{\label{fig:f4} \sl{Magnet geometric shapes.}} 
\end{figure}

Numerically, the magnetic field generated by a couple of big magnets is computed by dividing the large pieces into infinitesimal volumes, as shown in Fig.~\ref{fig:f4}. Using the corresponding magnetization per unit volume, it is possible to apply the dipole approximation between any two such small regions (one in each magnet), provided that their separation is much larger than their sizes. The final force is the sum of all these infinitesimal contributions. For instance, in Fig.\ref{fig:f4}, the vector $\vec{r}$ gives the position of an infinitesimal volume of the fixed magnet (with respect to the origin $O$ of the reference system, chosen at the center of the fixed magnet), while $\overrightarrow{r'}$ points on the levitated magnet. $\Delta \vec{r}$ is defined as the separation vector between the two considered volumes.

The computation of the forces between extended magnets has been implemented in MATLAB \cite{MATLAB}. The main aim of this tool is to study the best configuration, by changing the dimensions of the two magnets and their separation. Of course, the accuracy of the computation depends on the dimensions of the infinitesimal volume adopted for the simulation. In each computation, they have been reduced iteratively until convergence (within 2\%). In our numerical code, the shape, the size and the separation of the magnets are free parameters; for simplicity, we use parallelepipedal magnets.
%, i.e. when the relative difference between the previous value and the new one is below a few percents.

\subsection{Simulation results}
\label{subsec:simulation}

For a given geometrical configuration, the force between the magnets depends on the residual magnetization $B_r$, a magnet proper parameter. For our computation, we use Nd-Fe-B magnets which have currently the higher residual magnetization ($B_r=1.3$~T); the density of this material is $\rho=7.4~\mathrm{g/cm}^3$. For each magnet, the dimensions to be optimized are the length, that is the size of the rectangular parallelepiped in the direction the optical axis of the suspended mirror, the width, defined as the size along the transfer dimension and the vertical thickness.

The goal of our study is to determine how the static force between the fixed and the levitated magnet changes with their sizes and their separation. In particular, we use as parameter the `free-gap' $d$, i.e. the distance between the two magnets. In this way, we can evaluate the maximum mass that can be levitated in each configuration. In all cases, we need to subtract to the total levitated mass $M$ the weight of the levitated magnet $m$, so that we get the effective payload (mirror + marionetta) mass $M_p$.

Another subject studied here is the dependence of the mass to be suspended and of the horizontal restoring force on the length and the width of the magnets. This allows us to study the performance of the levitated system as a seismic isolator \cite{Drever}.

\begin{figure}
 \epsfxsize=19cm
 \centerline{\epsffile{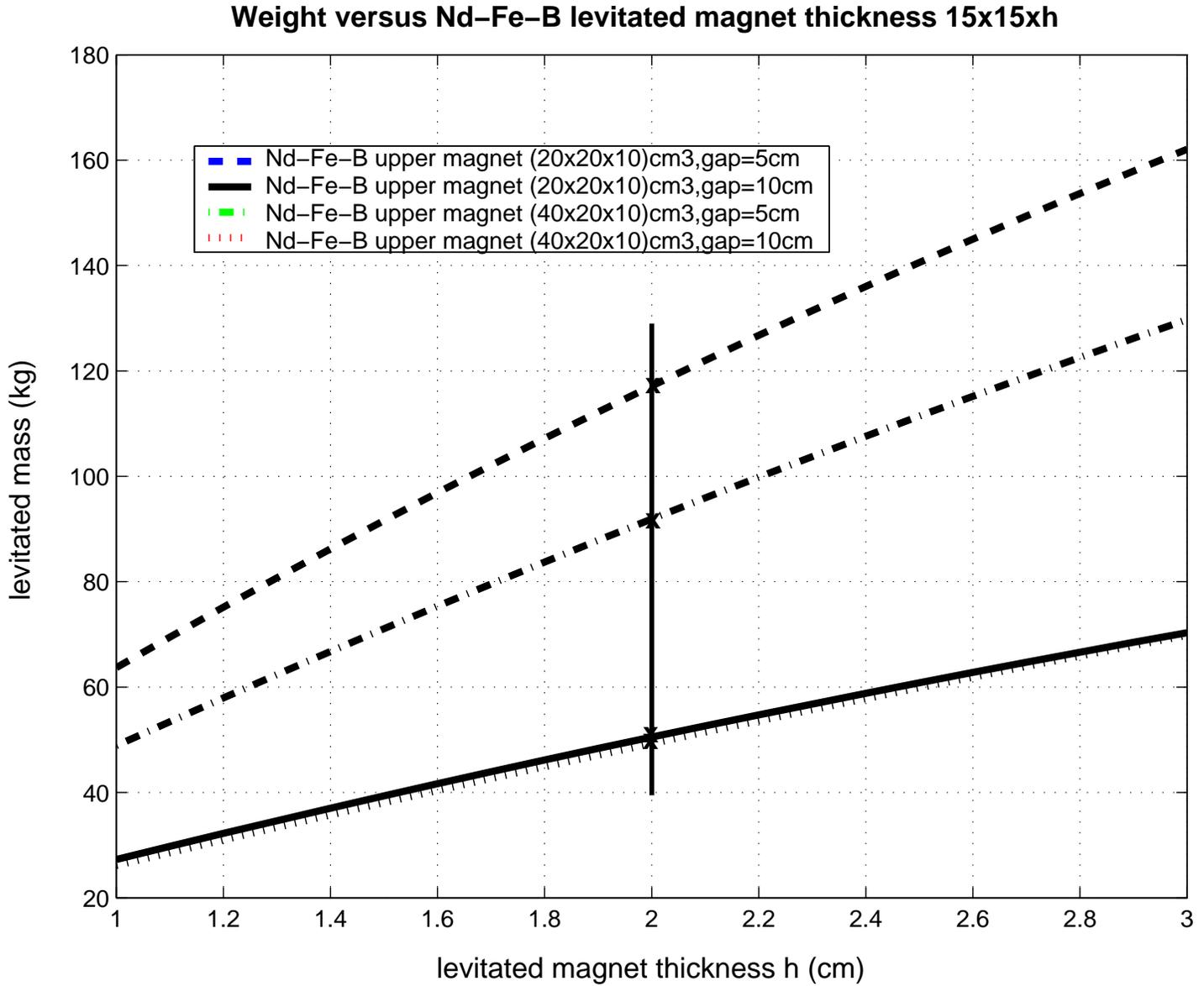}}
 \caption{\label{fig:f5} \sl{Variation of the maximum mass to be suspended versus the levitated magnet thickness $h$, for different gaps and different upper magnets -- in this plot, the levitated magnet weight has been already subtracted. The vertical line shows the weights computed at $h=2$~cm, which are used as reference results in the conclusion of the article.}}
\end{figure}

The first simulation aims at evaluating the maximum weight of the levitated mass $M_p$ as a function of the levitated magnet thickness {\sl{h}}. Fig.~\ref{fig:f5} shows the result: the simulation has been done for two different ``free-gap'' values ($d=5$~cm and $d=10$~cm) and for two different upper magnet configurations: $ (20\times 20\times 10) \ \mathrm{cm}^3$ and $(40\times 20\times 10) \ \mathrm{cm}^3$. The levitated magnet dimensions are $ (15\times 15\times h) \ \mathrm{cm}^3 $.

For instance, when the thickness of the levitated magnet to be $h=2$~cm, it is possible to suspend about $ 50$~kg when the free-gap is $10~cm$ and for upper magnet dimensions of $ (20 \times 20 \times 10) \ \mathrm{cm}^3 $. As we can see in  Fig.~\ref{fig:f5}, the maximum mass to be suspended increases with the thickness and of course  strongly depends on the free-gap. An interesting point, apparently counter-intuitive, is that for given values of $h$ and $d$, the force is smaller for the upper magnet with the larger lateral size. This aspect will be clarified later.

\begin{figure}
 \epsfxsize=19cm
 \centerline{\epsffile{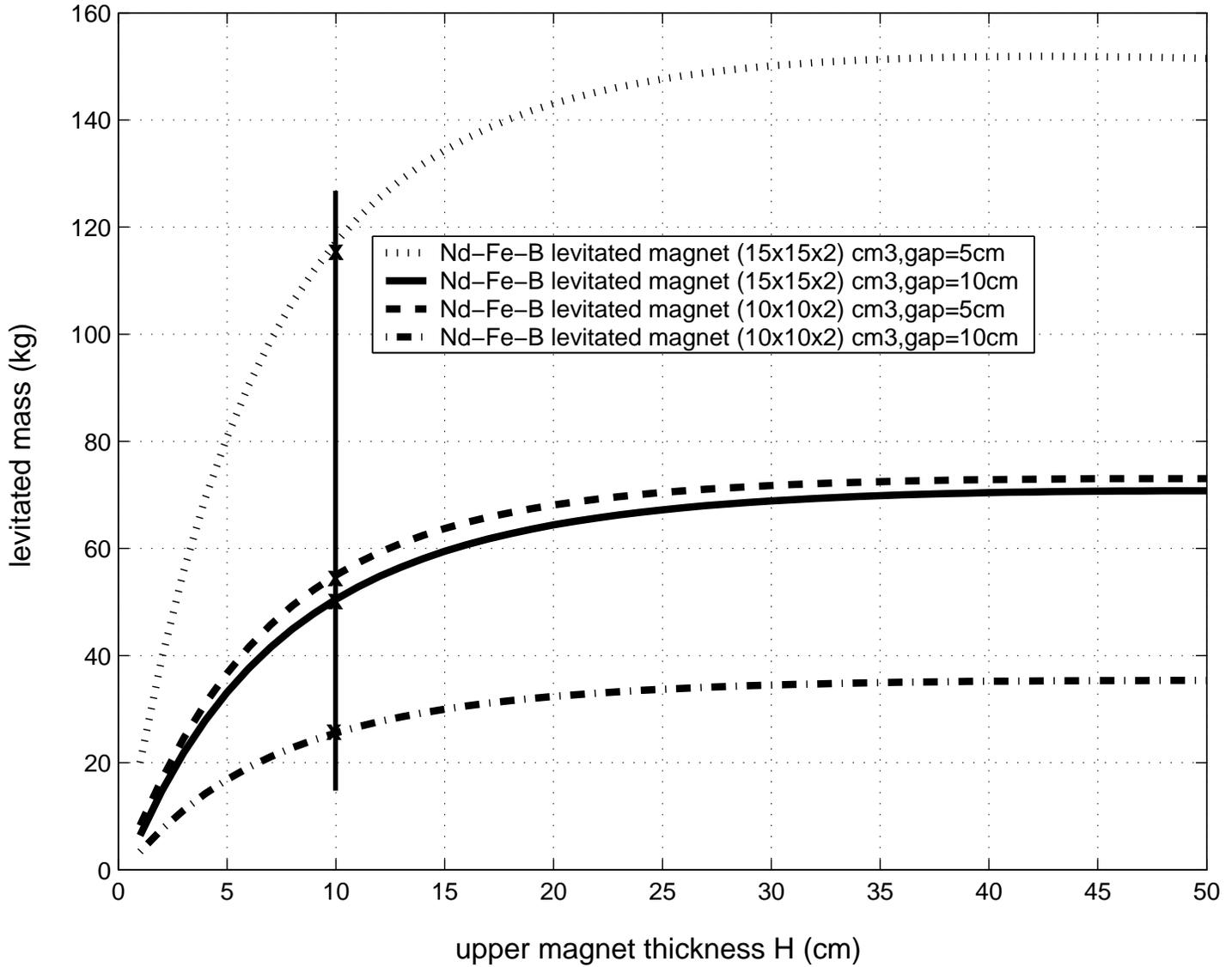}}
 \caption{\label{fig:f6} \sl{Variation of the maximum mass to be suspended versus the upper magnet thickness $H$ -- in this plot, the levitated magnet weight has been already subtracted. The vertical line shows the weights computed at $H=10$~cm, which are used as reference results in the conclusion of the article.}}
\end{figure}

The second step is to evaluate the dependence of $M_p$ on the upper magnet thickness {\sl{H}}. The result is shown in Fig.~\ref{fig:f6}: simulations have been done for two free-gap values already considered and for two different levitated magnet configurations: the first one, with dimensions $ (15\times 15\times 2) \ \mathrm{cm}^3 $ and mass $M_m \sim 3.3$~kg and the second one, with dimensions $ (10\times 10\times 2) \ \mathrm{cm}^3 $ and mass $M_m \sim 1.5$~kg; the upper magnet dimensions are $ (20\times 20\times H) \ \mathrm{cm}^3$. We can see that it is possible to obtain a levitated mass of $120.5$~kg in the configuration $(15\times 15\times 2) \ \mathrm{cm}^3 $ and with a gap of 5~cm: as a cross-check, this value corresponds to the one found in Fig.~\ref{fig:f5} with the same configuration.

Looking at Fig.~\ref{fig:f6}, we can see that it is almost useless to increase the thickness $H$ above $30 \div 40$~cm because the force becomes almost constant. As expected, there is still a strong dependence on the gap; for example, we obtain almost the same force with a much smaller suspended magnet, by reducing $d$ from 10~cm to 5~cm.

\begin{figure}
 \epsfxsize=19cm
 \centerline{\epsffile{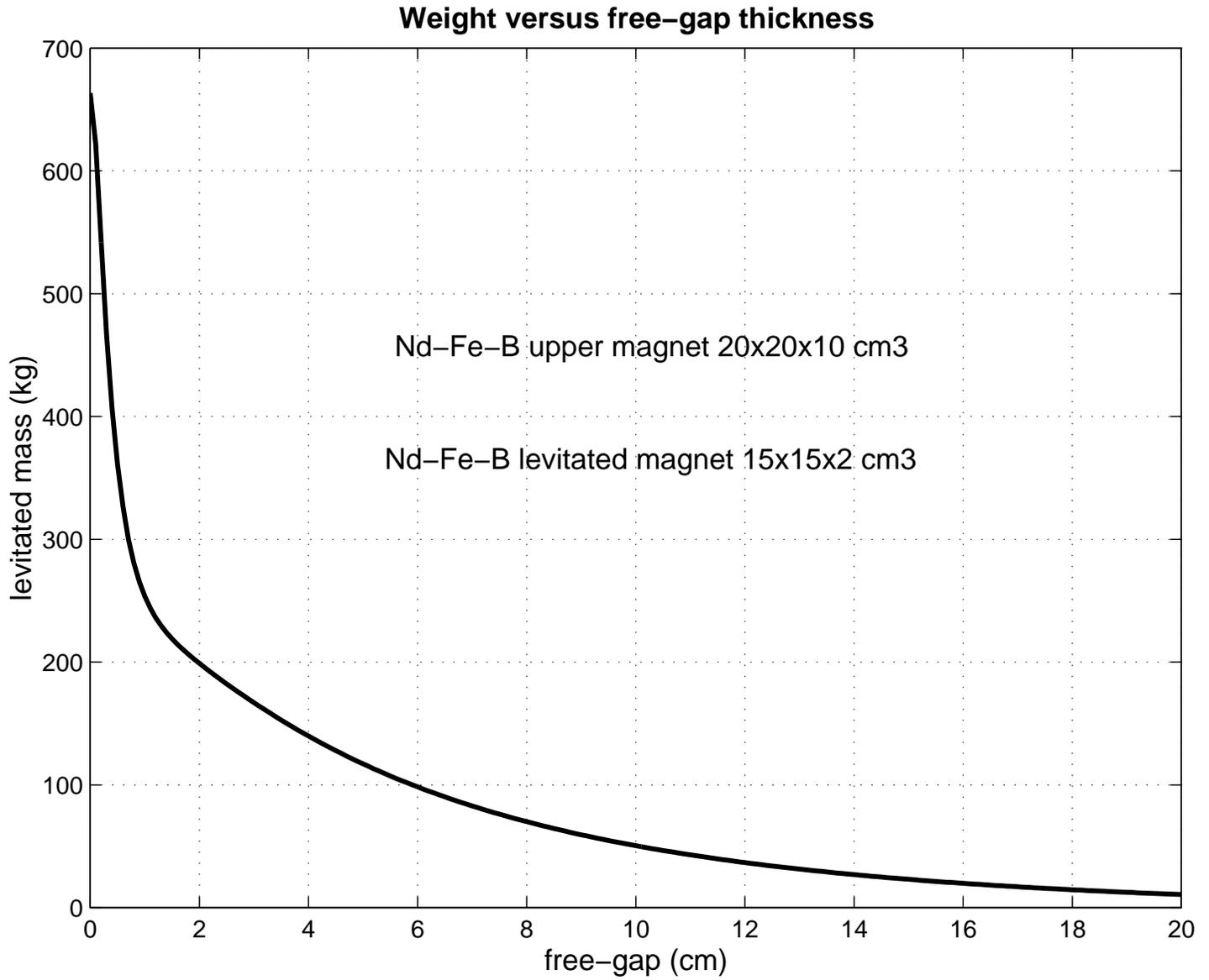}}
 \caption{\label{fig:f7} \sl{Maximum mass to be suspended versus the free-gap thickness with a $ 20\times 20\times 10 \ \mathrm{cm}^3 $ upper magnet and a $ 15 \times 15\times 2 \ \mathrm{cm}^3 $ levitated magnet -- in this plot, the levitated magnet weight has been already subtracted.}}
\end{figure}

The third simulation studies the evolution of the levitated mass $M_p$ when the free-gap $d$ varies, as Fig.~\ref{fig:f7} shows; the upper magnet dimensions are, in this case, $ (20\times 20\times 10) \ \mathrm{cm}^3 $ and the levitated ones $ (15\times 15\times 2) \ \mathrm{cm}^3 $. Of course, the suspendable mass value decreases with the increase of the free-gap.

\begin{figure}
 \epsfxsize=19cm
 \centerline{\epsffile{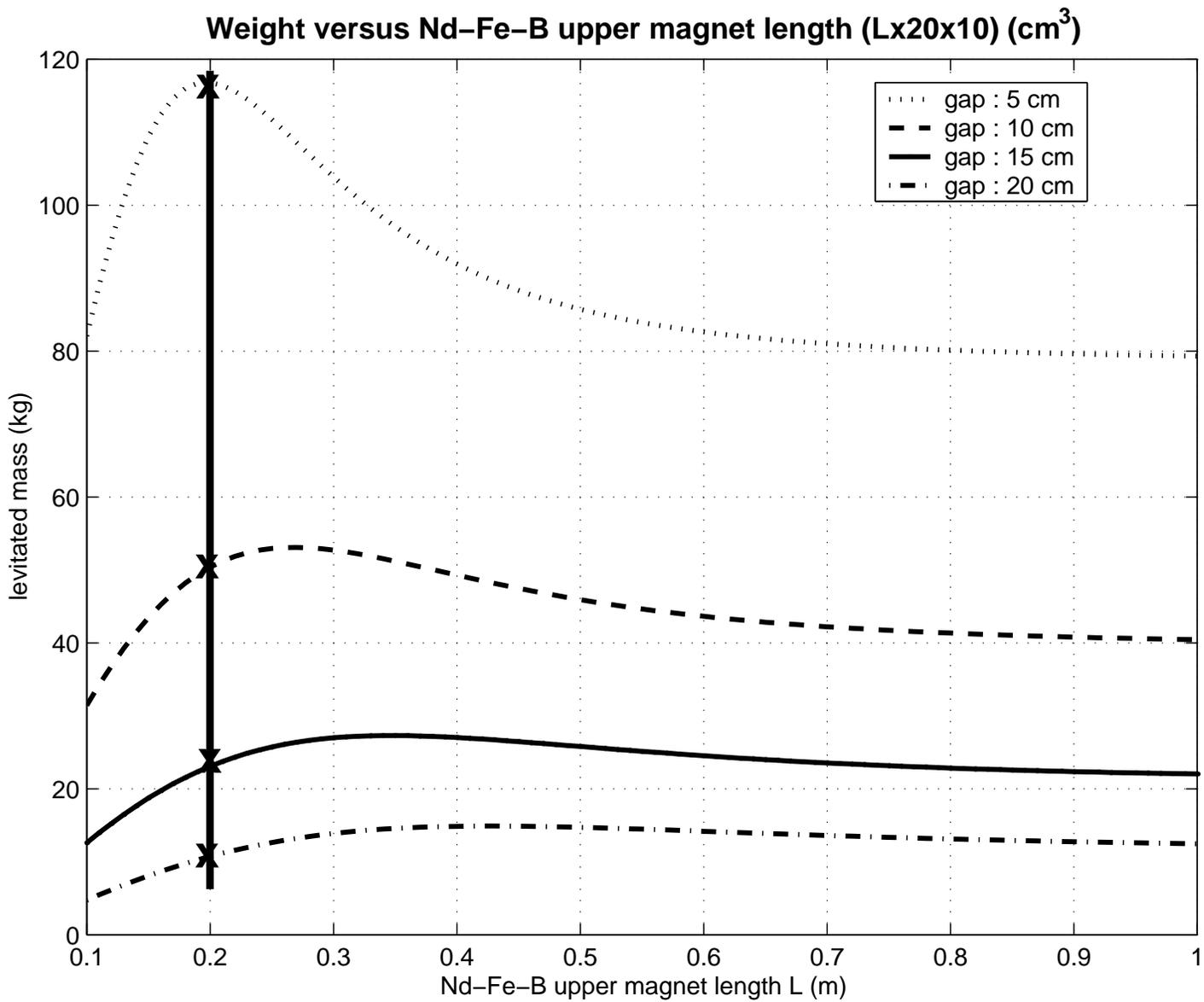}}
 \caption{\label{fig:f8} \sl{Variation of the maximum mass to be suspended versus the upper magnet length $L$ -- in this plot, the levitated magnet weight has been already subtracted.}}
\end{figure}

The variation of the suspendable mass $M_p$ versus the length {\sl{L}} of the upper magnet is the topic of the fourth simulation. As shown in Fig.~\ref{fig:f8}, the calculation has been done for a set of different free-gap values ranging from $d=5$~cm to $d=20$~cm. The dimensions of the magnets are $ (L\times 20\times 10) \ \mathrm{cm}^3 $ for the upper one and $ (15\times 15\times 2) \ \mathrm{cm}^3 $ for the levitated one. As we can see from the Figure, using an upper magnet too long with respect to the free-gap dimension is not an advantage, because the force crosses a maximum and then decreases asymptotically to a constant value. This effect has been already observed in Fig~\ref{fig:f5}. It can be easily explained by noting that the force along the magnetic dipole direction changes its sign when the transverse distance of the two dipoles becomes much larger than their longitudinal one.

\begin{figure}
 \epsfxsize=19cm
 \centerline{\epsffile{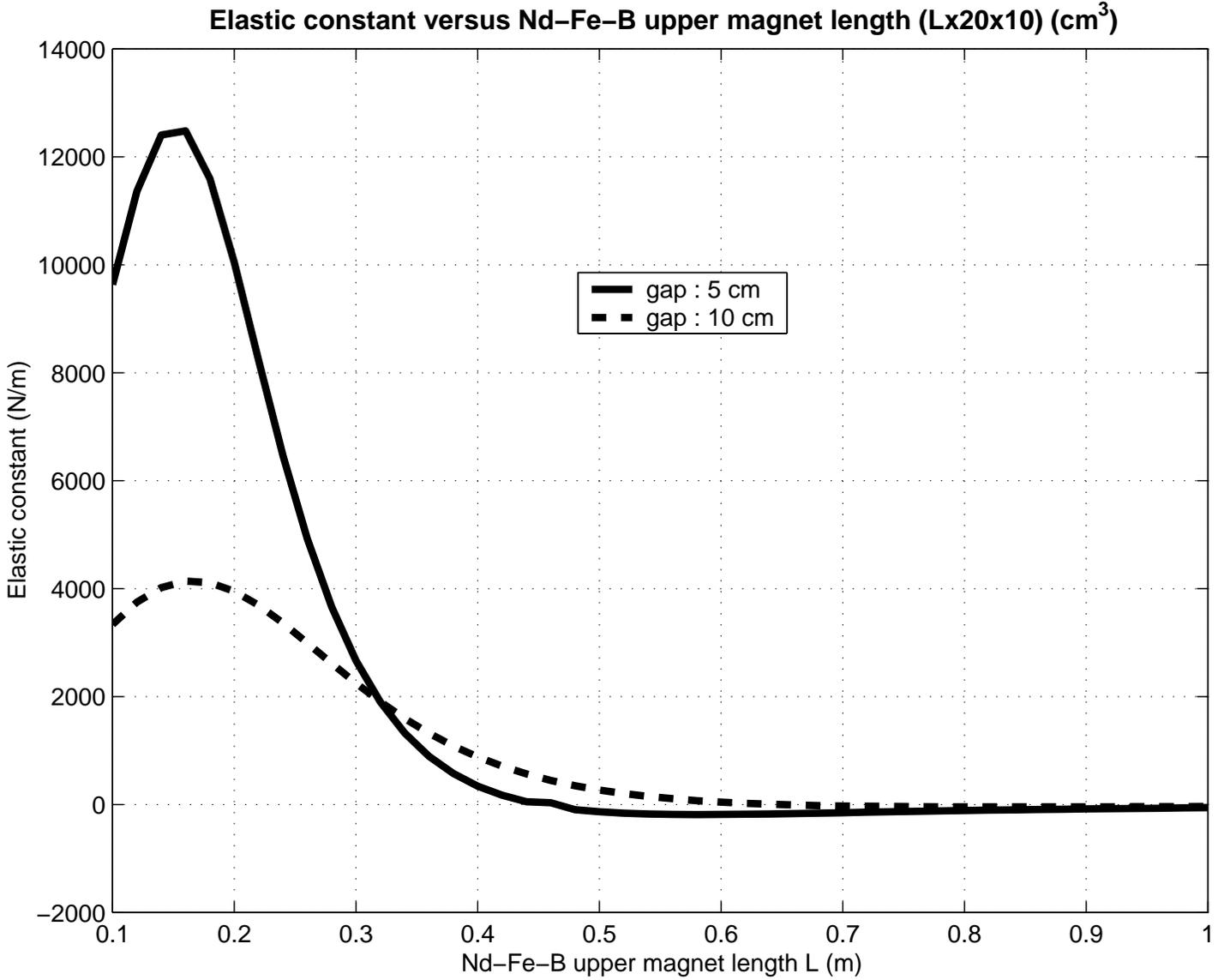}}
 \caption{\label{fig:f9} \sl{Dependence of the restoring force along the mirror optical axis direction on the variation of the upper magnet length L: results are shown for two different free-gap values ($d=5$~cm and $d=10$~cm), for a levitated magnet $ (15\times 15\times 2) \ \mathrm{cm}^3 $ and for a constant misalignment of 1~mm along the length direction.The restoring force is always vanishing for $L \longrightarrow \infty$, but for some gap-length configurations, it can become negative: in this case the system gets unstable in the horizontal direction, i.e. for $d=5$~cm the restoring force is equal to zero, ranging $40 \div 50$~cm; so, for $L$ approaching this value, we can get a stable configuration with very small restoring force even with a finite magnet.}}
\end{figure}

The last investigated point is the dependence of the restoring force along the mirror optical axis direction on the variation of the upper magnet length {\sl{L}}. This calculation has been done for different free-gap values, $d=5$~cm and $d=20$~cm, for a levitated magnet $ (15\times 15\times 2) \ \mathrm{cm}^3 $ and for a constant `misalignment' of 1~mm along the length direction, as shown in Fig.~\ref{fig:f9}. As we can see, the restoring force is always vanishing for $L \longrightarrow \infty$; in this case, as proposed by Drever \cite{Drever}, the system behaves like a pendulum with a very low resonant frequency, i.e. it is in principle a very good seismic isolator.

The interesting point is that, for some gap-length configurations, the restoring force can become negative: in this case the system gets unstable in the horizontal direction. As a consequence, there is a finite length giving a restoring force equal to zero, ranging $40 \div 50$~cm. So, for $L$ approaching this value, we can get a stable configuration with very small restoring force even with a finite magnet. For example, if we take a magnet with $L=40$~cm, with $d=5$~cm we can levitate up to $\sim 95$~kg (see Fig.~\ref{fig:f9}) with an horizontal restoring force of only 200~N/m, corresponding to a resonant frequency of $\sim 0.23$~Hz: this is the equivalent of a 45~m-long pendulum. As a comparison, the SA main resonance frequency is around 30~mHz which is equivalent to a 275~m-long simple pendulum.

\subsection{Experimental test}
\label{subsec:test}
 
\begin{figure}
 \epsfxsize=19cm
 \centerline{\epsffile{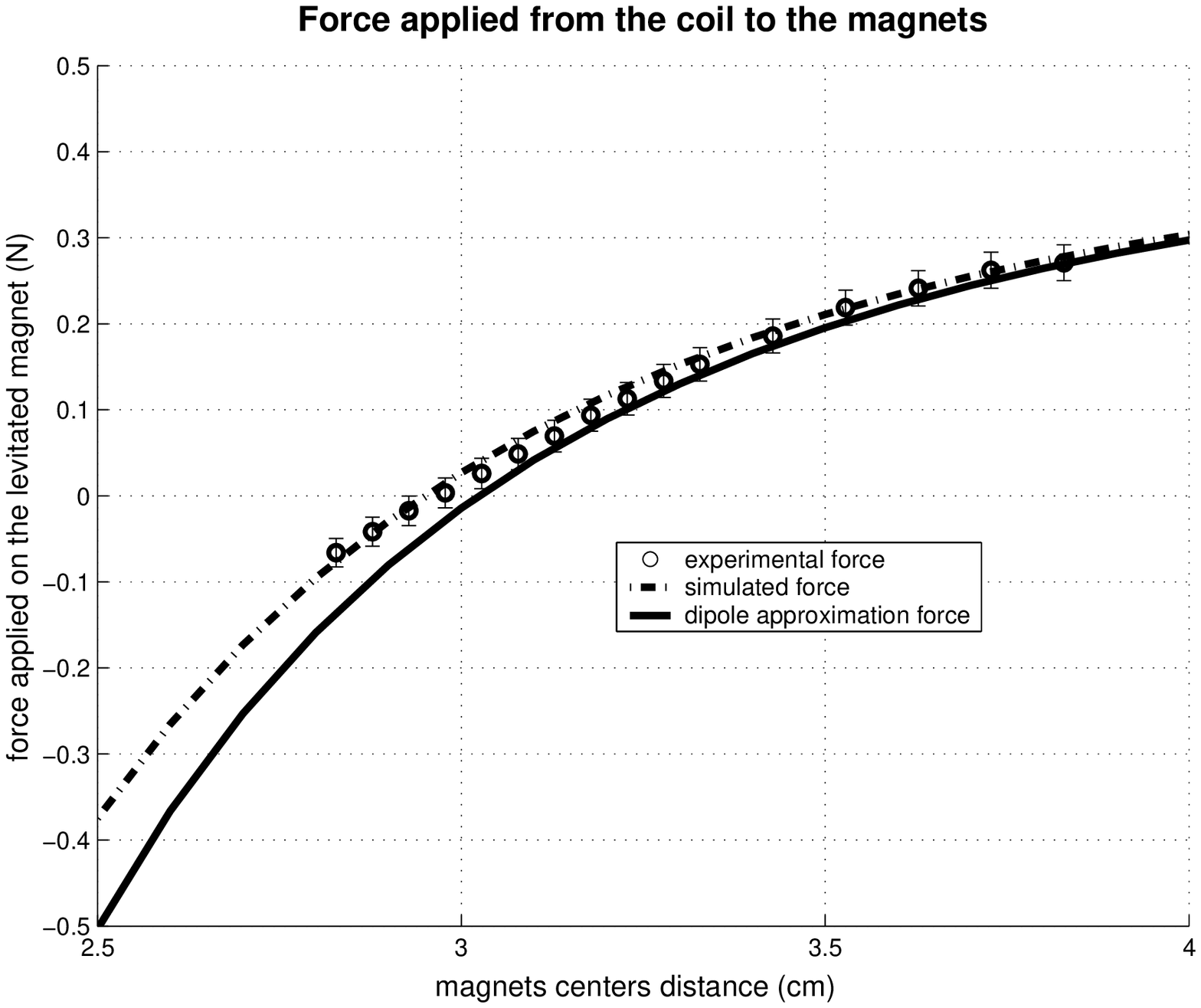}}
 \caption{\label{fig:f10} \sl{Dependence of the force measured between the magnets on the distance between the magnet centers. The theoretical curve computed with our simulation model (dashed line) has been superimposed to the experimental points which are in agreement with the model. For comparison, the force computed by approximating the extended magnets as point-like dipoles placed in their centers is added; as expected the point-like dipole approximation becomes unsatisfactory as the distance decreases and is in disagreement with the experiment for a distance below 3~cm.}}
\end{figure}

A magnetic suspension prototype has been realized in Naples \cite{Varvella} to check the correctness of our models and to verify that we take into account all the relevant effects. Our set-up is sketched in Fig.~\ref{fig:f3}; we use {\sl Sm-Co} cylindrical magnets which have a residual magnetization $B_r=0.8$~T and a density $\rho=8.3~\mathrm{g/cm}^3$. The radius is $R = 7$~mm while the thickness is $h = 12$~mm for the fixed magnet and 8~mm for the levitated one with a mass $m$ of about 10~g. To measure the vertical position of the payload, we use a shadow-meter sensor made of a laser diode partially intersected by the payload lower edge and a photodetector. To get a stable position we feedback on a coil acting on the levitated magnet with a force/current characteristic of about 1~N/A. In this way we are able to hold constant the vertical position of the payload respect to the ground. Using this configuration it is possible to suspend a payload of about 45~g with a free-gap ranging between 1 and 3~cm. The `fixed' magnet is mounted on a micrometric translator allowing to change the free-gap and consequently the vertical force between the magnets. When this force exactly balanced the total weight of the suspended body, the DC current flowing in the coil is zero. The changing of the vertical distance of the magnets with the micrometer results in a non-zero DC current because the force exerted by the coil must balance the difference between the weight and magnets forces. In this way, the current flowing in the coil provides a measurement of the force between the magnets (subtracted of the weight of the suspended mass). Fig.~\ref{fig:f10} shows the dependence of the measured force between the magnets on the distance between the centers of the magnets. The theoretical curve computed with our simulation model (dashed line) is superimposed to the experimental points. As we can see, the experimental points are in agreement with the model. For comparison, we add the force computed by approximating the extended magnets as point-like dipoles placed in their centers; as expected, the point-like dipole approximation becomes unsatisfactory as the distance decreases and disagrees with the experiment for a distance below 3~cm.

\section{Conclusion}
\label{sec:conclusion}

In this paper we studied the application of a magnetic levitation system to the seismic suspensions of long baseline GW antennas. From the results of the simulations shown in Sec.~\ref{subsec:simulation}, we can summarize the following conclusions.

First, it appears that it is possible to levitate a mass comparable to the one of a Virgo-like payload, choosing reasonable dimensions for both the fixed and levitated magnet. This can also be done maintaining a free-gap of several centimeters, which would allow the insertion of a dielectric roof to separate the upper and the lower vacuum chambers.

A second interesting point is that, with a suitable magnet arrangement, the levitated system behaves also as a low frequency seismic attenuator, which improves thus the overall attenuation performance of the suspension. To improve the stability of the device, a pair of magnets aligned with opposing polarity can be used instead of a single one, as first suggested in Ref. \cite{Drever}. This configuration was validated experimentally in the Naples Virgo laboratory \cite{Varvella}.

To give an example, we can consider as convenient configuration a fixed magnet of $ (40\times 20 \times 10) \ \mathrm{cm}^3 $, a levitated one of $ (15\times 15\times 2) \ \mathrm{cm}^3 $ and a gap $d=5$~cm. In this configuration, the weights of the two magnets are 66.4~kg and 3.3~kg respectively, while the levitated payload $M_p$ is $\sim 95$~kg. As explained before, the horizontal oscillation frequency is, in this case, 0.23~Hz, giving an extra attenuation of about 74~db at 10~Hz.

The validity of the numerical model adopted for computing forces exerted between extended magnets has been experimentally tested with a small prototype suspension. The experimental results are in good agreement with the model. Of course, the study presented here only shows the feasibility of the principle; one of the main limitation is that we did not consider the coupling with the angular degrees of freedom of the suspended payload. A detailed study would be necessary to design a real suspension taking into account the need of controlling the other degrees of freedom of the mirror and other technical aspects, like the longitudinal control and the automatic alignment of the interferometer. In addition, a suspension designed for GW interferometers must not inject too many noise in the detector. Therefore, noise contributions originating from the magnetic configuration itself -- like eddy and Johnson current effects -- have to be taken into account on a full-scale prototype to see whether or not they limit the suspension performances. This is beyond the goal of this paper.

\newpage

\bibliographystyle{plain}

\newpage

\begin{table}[ht!]
 \begin{center}
  \caption{\label{tab:t1} \sl{Masses of the SA components.}}
  \vspace{2mm}
  \begin{tabular}{|c|c|} \hline
   Component               & Mass (kg) \\[-0.5mm] \hline
   SA                      & 1000      \\[-0.5mm]
   {\sl{Marionetta}}       &  80       \\[-0.5mm]
   Mirror                  &  20       \\[-0.5mm] 
   {\sl{Reference mass}}   &  20       \\         \hline
  \end{tabular}
 \end{center}
\end{table}

\newpage

\newpage

\newpage

\newpage

\newpage

\newpage

\newpage

\newpage

\newpage

\newpage

\end{document}